\documentclass{aa}

\pdfoutput=1

\usepackage{graphicx}
\usepackage{txfonts}
\usepackage[colorlinks=true,urlcolor=blue,citecolor=blue]{hyperref}

\begin{document}

\title{MASCARA-1\,b}
\subtitle{A hot Jupiter transiting a bright $m_V=8.3$ A-star in a misaligned orbit}
\author{G.J.J. Talens \inst{1}
\and S. Albrecht \inst{2}
\and J.F.P. Spronck \inst{1}
\and A.-L. Lesage \inst{1}
\and G.P.P.L. Otten \inst{1}
\and R. Stuik \inst{1}
\and V. Van Eylen \inst{1}
\and H. Van Winckel \inst{4}
\and D. Pollacco \inst{3}
\and J. McCormac \inst{3}
\and F. Grundahl \inst{2}
\and M. Fredslund Andersen \inst{2}
\and V. Antoci \inst{2}
\and I.A.G Snellen \inst{1}}

\institute{Leiden Observatory, Leiden University, Postbus 9513, 2300 RA, Leiden, The Netherlands\\
\email{talens@strw.leidenuniv.nl} \and Stellar Astrophysics Centre (SAC), Department of Physics and Astronomy, Aarhus University, Ny Munkegade 120, DK-8000 \and Department of Physics, University of Warwick, Coventry CV4 7AL, UK \and Instituut voor Sterrenkunde, KU Leuven, Celestijnenlaan 200D bus 2401, 3001 Leuven, Belgium}

\abstract{We report the discovery of MASCARA-1\,b, the first exoplanet discovered with the Multi-site All-Sky CAmeRA (MASCARA). It is a hot Jupiter orbiting a bright $m_V=8.3$, rapidly rotating ($v\sin i_\star > 100~\rm{km~s}^{-1}$) A8 star with a period of $2.148780\pm8\times10^{-6} ~\rm{days}$. The planet has a mass and radius of $3.7\pm0.9~\rm{M}_{\rm{Jup}}$ and $1.5\pm0.3~\rm{R}_{\rm{Jup}}$, respectively. As with most hot Jupiters transiting early-type stars we find a misalignment between the planet orbital axis and the stellar spin axis, which may be signature of the formation and migration histories of this family of planets. MASCARA-1\,b has a mean density of $1.5\pm0.9~\rm{g~cm^{-3}}$ and an equilibrium temperature of $2570^{+50}_{-30}~\rm{K}$, one of the highest temperatures known for a hot Jupiter to date. The system is reminiscent of WASP-33, but the host star lacks apparent delta-scuti variations, making the planet an ideal target for atmospheric characterization. We expect this to be the first of a series of hot Jupiters transiting bright early-type stars that will be discovered by MASCARA.}

\maketitle

\section{Introduction}
\label{sec:introduction}

Since the discovery of the first exoplanet by \citet{Mayor1995} the number of new planets detected has been growing exponentially. First from ground-based radial velocity \citep{Vogt2000,Valenti2005,Jenkins2009} and transit surveys \citep{Bakos2004,Pollacco2006} and in recent years from the CoRoT and \emph{Kepler} satellites \citep{Barge2008,Borucki2010}. Today, several thousand exoplanets are known and our understanding of the population has grown immensely. In particular, we now know that on average a late-type main-sequence star is orbited by at least one planet \citep{Batalha2014}. 

Despite these advances, observational biases mean we still know only a few planets transiting both bright and early-type stars. The Multi-site All-Sky CAmeRA \citep[MASCARA,][]{Talens2017} transit survey aims to fill this gap in parameter space by monitoring all stars with magnitudes $4 < m_V < 8.4$. MASCARA consists of a northern and a southern station, each equipped with five cameras to cover the entire local sky down to airmass 2 and partially down to airmass 3. In this way MASCARA is expected to find several new transiting hot Jupiters around bright stars, some of which will be of early spectral types \citep{Snellen2012}. 

MASCARA was designed to find Jupiter-like planets transiting bright stars because these allow detailed atmospheric characterization. Observations of the planet transmission spectra during transit and thermal emission at secondary eclipse, coupled with detailed atmospheric models, can be used to constrain atmospheric properties such as temperature structure, composition and the presence of clouds and hazes \citep{Madhusudhan2009,Sing2011,Deming2013,Kreidberg2014,Sing2016}. Such observations, however, are expensive if not impossible on fainter stars. In particular, the use of high-resolution transmission spectroscopy has been limited to a few bright targets such as HD\,209458 and HD\,189733 \citep{Snellen2010,Brogi2016} which were first discovered trough radial velocity observations.

In addition to finding planets around bright stars, the all-sky, magnitude-limited sample targeted by MASCARA is biased towards early type stars, which are particularly challenging to follow-up at fainter magnitudes. Studying the hot Jupiter population around hot stars is interesting for several reasons. First, these hot Jupiters will have the highest equilibrium temperatures, which is expected to have a significant impact on their atmospheric structure and composition - such as the occurrence of inversion layers due to the presence of specific molecular compounds \citep[e.g. TiO/VO][]{Hubeny2003,Fortney2008} high up in the atmospheres (e.g. \cite{Burrows2007,ODonovan2010} but see also \citet{Schwarz2015}). Second, hot Jupiters around early type stars receive more UV radiation than similar planets around late type stars. This UV radiation may drive unique chemical processes in the atmospheres of these planets \citep{Casewell2015}. Finally, comparing their orbital properties with those of hot Jupiters orbiting solar-type stars may shed light on their formation. For example, the large incidence of misaligned orbits for hot Jupiters orbiting early-type stars \citep{Winn2010,Schlaufman2010,Albrecht2012} may be linked to orbital migration processes \citep[e.g. ][]{Fabrycky2007,Nagasawa2008}.

In this paper we present the discovery of MASCARA-1\,b, the first exoplanet discovered by MASCARA. MASCARA-1\,b orbits the $m_V=8.3$ A8 star HD\,201585 with a period of $2.15~\rm{days}$. We present our observations in Section \ref{sec:observations}. Stellar and system parameters are described in Sects. \ref{sec:star} and \ref{sec:system}, and we conclude with a discussion in Sect. \ref{sec:discussion}. 

\section{Observations}
\label{sec:observations}

MASCARA's northern station is located on La Palma and started science operations early 2015, producing photometry for all stars with $4 < m_V < 8.4$ at a cadence of 6.4 seconds down to airmass $\sim$3 \citep{Talens2017}. The raw MASCARA photometry is processed using a heavily modified version of the coarse decorrelation algorithm described in \citet{CollierCameron2006} in order to remove instrumental systematics and earth-atmospheric effects. Subsequently, the data are binned to a cadence of 320 seconds and further detrended by means of an empirical fit to remove both residual variations linked to the target positions on the CCDs and long-term trends in the data. Finally, a Box Least-Squares \citep[BLS,][]{Kovacs2002} transit search algorithm is used to find potential transit signals. The details of the data reduction and transit search algorithms will be discussed in an upcoming paper (Talens et al. in prep).

\begin{table}
\centering
\caption{Observations used in the discovery of MASCARA-1\,b.}
\begin{tabular}{l c c c}
 Instrument & Date & $N_{\rm{obs}}$ & $t_{\rm{exp}}$ [s] \\ 
 \hline
 \hline
 MASCARA & Feb. 2015 - Sept. 2016 & 32705 & 320 \\  
 HERMES & 15 June 2016 & 1 & 1400 \\
 HERMES & 17 July 2016 & 1 & 800\\
 HERMES & 18 July 2016 & 1 & 800\\
 NITES & 23 July 2016 & 1092 & 10 \\
 HERMES & 5 Sept. 2016 & 1 & 1100\\
 HERMES & 7 Sept. 2016 & 1 & 1200\\
 HERMES & 7 Sept. 2016 & 1 & 1100\\
 HERMES & 9 Sept. 2016 & 1 & 1100\\
 HERMES & 10 Sept. 2016 & 1 & 1000\\
 HERMES & 10 Sept. 2016 & 1 & 1200 \\
 HERMES & 11 Sept. 2016 & 1 & 1000\\
 HERMES & 12 Sept. 2016 & 1 & 1200 \\
 HERMES & 13 Sept. 2016 & 1 & 900 \\
 HERMES & 14 Sept. 2016 & 1 & 1100\\
 NITES & 17 Sept. 2016 & 887 & 15 \\
 SONG & 30 Sept. 2016 & 28 & 600
\end{tabular}
\label{tab:datasets}
\end{table}

After running the BLS algorithm on data taken  between February and December 2015, MASCARA-1, (HD\,201585, HIP\,104513) was among the most promising stars to be listed as a potential exoplanet host. We obtained high-dispersion spectra at R=85,000 from the HERMES spectograph \citep{Raskin2011} at the 1.2m Mercator telescope on La Palma taken between June and September 2016 (see Table \ref{tab:datasets}). A total of 13 spectra were taken and reduced using the HERMES pipeline\footnote{\label{note:hermes}\url{http://hermes-as.oma.be/manuals/cookbook6.0.pdf}}, allowing us to characterize the host star and constrain the companion mass from radial velocity (RV) measurements.

In addition to the MASCARA photometry, two transits were observed with the 0.4m NITES telescope \citep{McCormac2014} on La Palma. 1,092 images with 10s exposure times were obtained with a Johnson-Bessel R-band filter on July 23, 2016 and 887 images with 15s exposure times were obtained in the V-band on September 17, 2016. The telescope was defocused slightly on both nights to avoid saturation. The data were reduced in Python with CCDPROC \citep{Craig2015} using a master bias, dark and flat. Master calibrations were made using a minimum of 21 of each frame. Non-variable comparison stars were selected by hand and aperture photometry extracted using SEP \citep{Barbary2016,Bertin1996}. The shift between each defocused image was measured using the DONUTS \citep{McCormac2013} algorithm and the photometry apertures were recentered between frames. The final aperture size is chosen to minimise the RMS of the data points out of transit. The $0.66\arcsec$ pixel scale of NITES, versus $1\arcmin$ for MASCARA, allowed us to check against a faint eclipsing binary system within the MASCARA photometric aperture that could have explained the apparent transit signal.

Additional confirmation about the nature of the system, as well as a measurement of its spin-orbit angle, can come from a detection of the Rossiter-McLaughlin effect using in-transit spectroscopy \citep[e.g.][]{Zhou2016}. We therefore obtained 28 high-dispersion echelle spectra at R$\sim$90,000 using the automated 1m Hertzsprung SONG telescope \citep{Andersen2014} at Observatorio del Teide on September 30, 2016 (see Table \ref{tab:datasets}). The observations were taken with 10 minute exposure times and a slit width of 1.2 arcsec and spanned ${\sim}5$~hr in total, covering the transit and post-egress. Ingress occurred during evening twilight. Spectra were extracted from the observations following the procedure outlined in \citet{Grundahl2017}, subsequently bad pixels were removed and the spectra were normalized. 

\begin{figure}
  \centering
  \includegraphics[width=8.5cm]{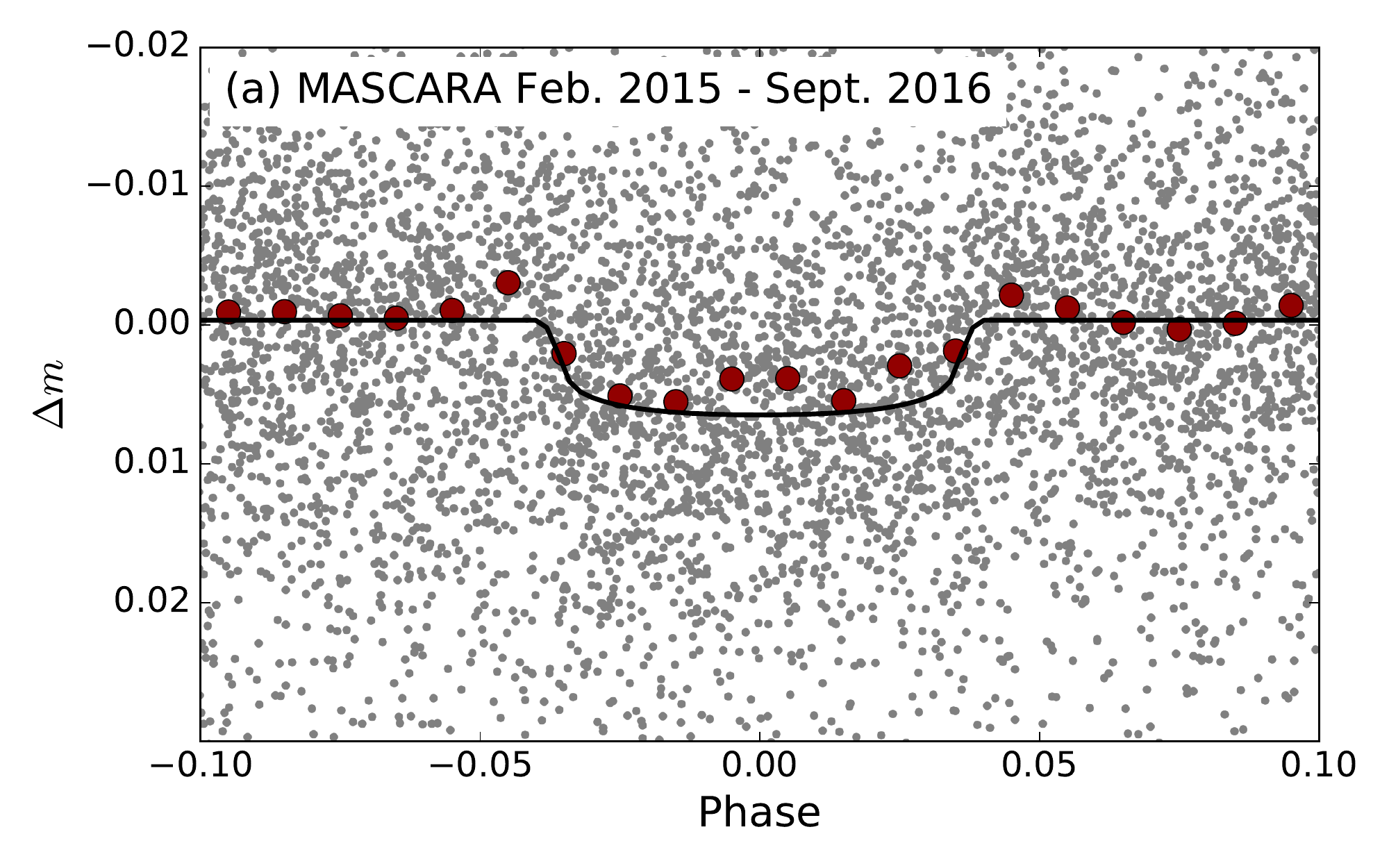}
  \includegraphics[width=8.5cm]{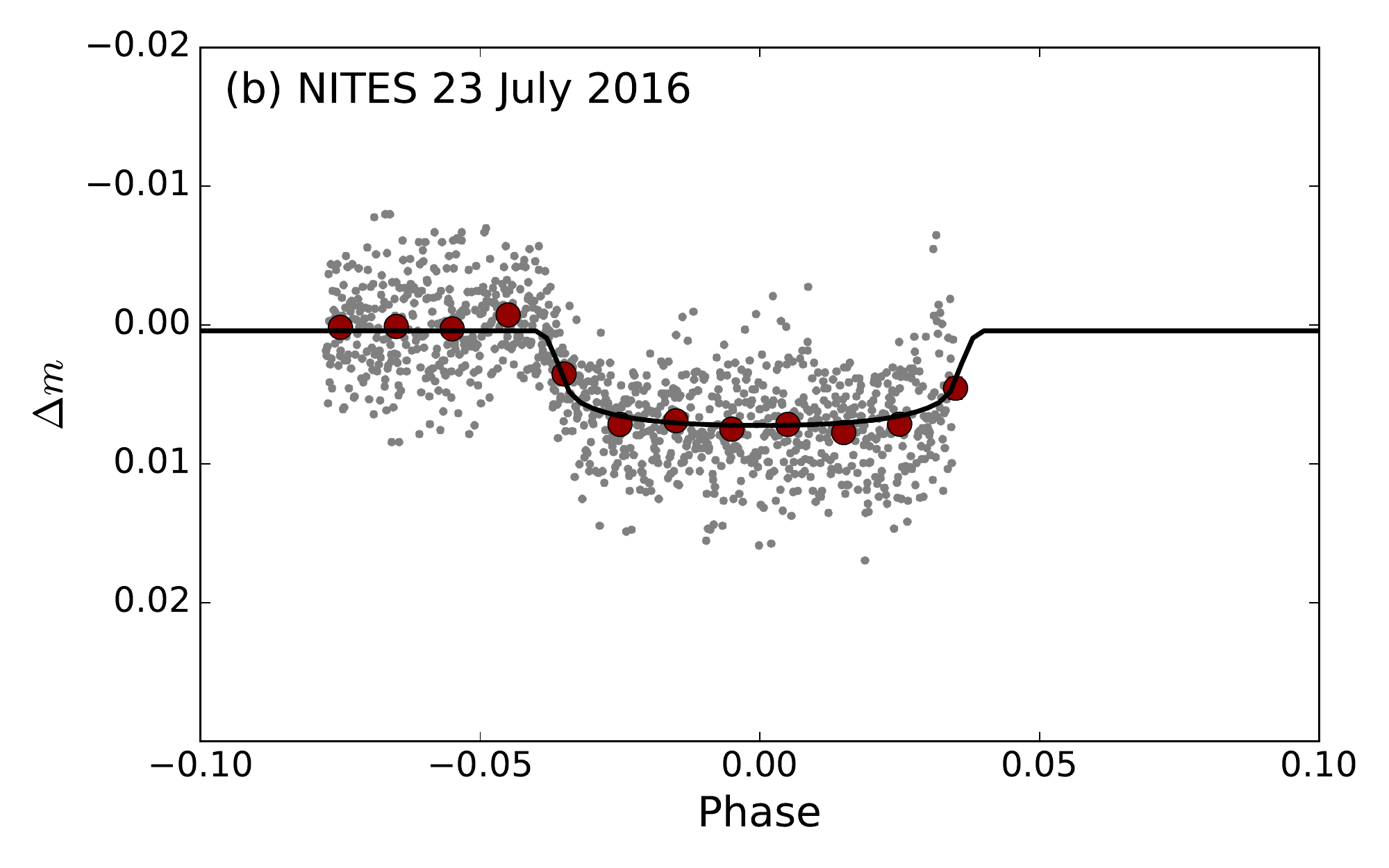}
  \includegraphics[width=8.5cm]{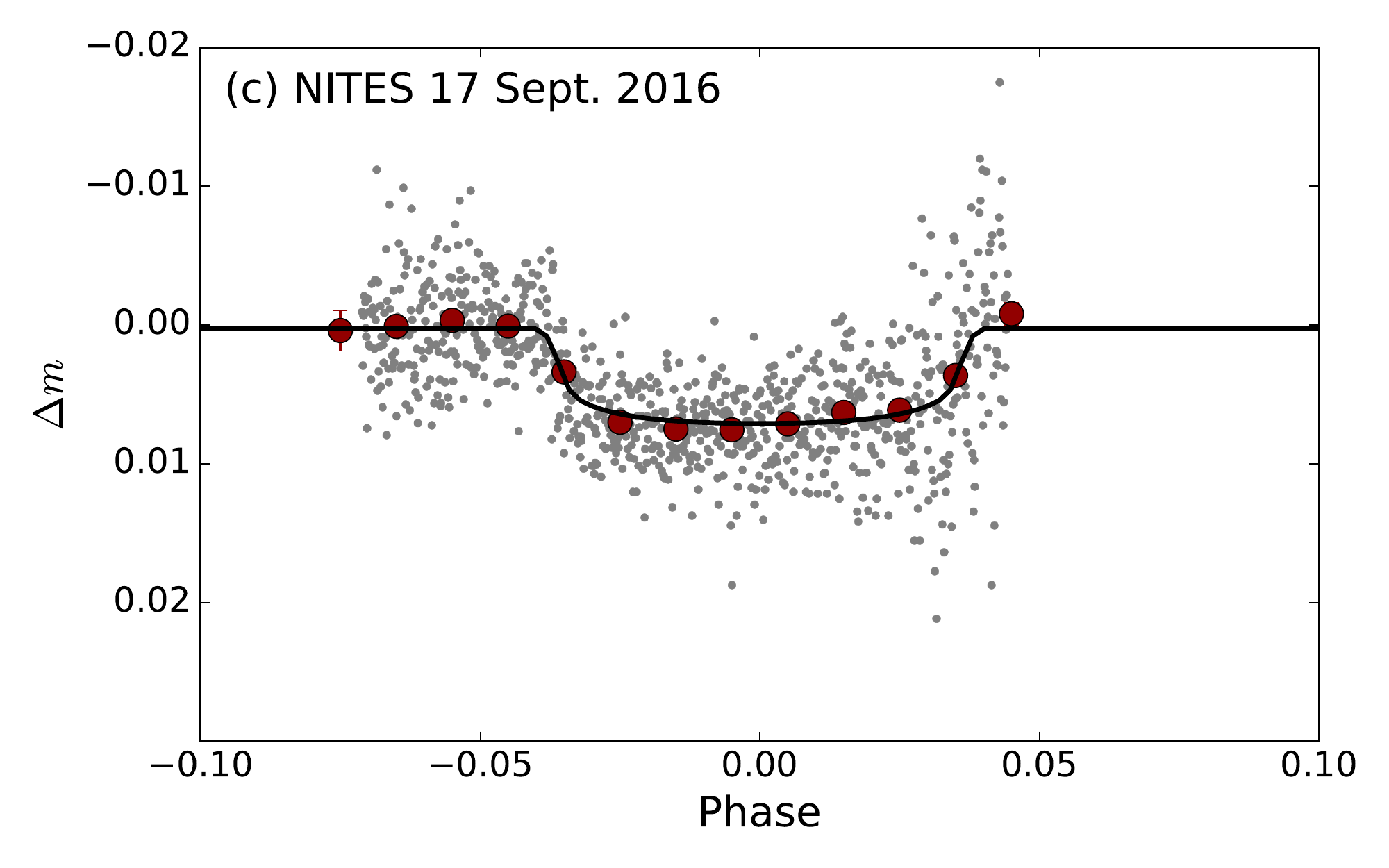}
  \caption{Photometry and best-fit models used in the characterization of MASCARA-1\,b. (a) MASCARA photometry obtained between February 2015 and September 2016. Shown are the 320 s cadence data (grey points), the data binned by 0.01 in phase (red points) and the best-fit transit model (black line). (b) and (c) NITES transit observations taken on 23 July 2016 and 17 September 2016, respectively. Shown are the data (grey points), the data binned by 0.01 in phase (red points) and the best-fit transit model (black line).}
  \label{fig:photometry}
\end{figure}

\section{Stellar parameters}
\label{sec:star}

At $m_V=8.3$, most basic properties of MASCARA-1 are available in the literature. The parallax was already measured by Hipparcos \citep{VanLeeuwen2007} to be $5.0\pm0.7~\rm{mas}$, implying a distance of $201.6\pm28.9~\rm{pc}$. Recently, the GAIA DR1 \citep{GAIA2016a} lists a parallax of $5.3\pm0.3~\rm{mas}$ ($188.7\pm10.7~\rm{pc}$), within 1$\sigma$ of the Hipparcos value. For our analysis we adopt the GAIA parallax.

\citet{McDonald2012} combined the Hipparcos parallaxes with photometry from a number of surveys, including SDSS and 2MASS, to determine the effective temperature and luminosity of over 100,000 stars, including MASCARA-1 for which they derive an effective temperature of $7554~\rm{K}$ and a bolometric luminosity of $14.98~\rm{L}_\odot$. We verified the effective temperature by fitting the wings of the hydrogen lines in the combined HERMES spectra with PHOENIX spectral models \citep{Husser2013} and find $T_{\rm{eff}} = 7600\pm150~\rm{K}$, consistent with the work done by \citet{McDonald2012}. For our further analysis, we used the values as derived by \citet{McDonald2012}, assuming uncertainties of $150~\rm{K}$ and $5~\rm{L}_\odot$ on $T_{\rm{eff}}$ and $L_*$ respectively\footnote{For $T_{\rm{eff}}$ we assume an uncertainty comparable to our own analysis. For $L_*$ the error on the distance alone implies an uncertainty of $4.3~\rm{L}_\odot$, which we round up to account for uncertainties in the total flux.} - which are not stated by the authors. We updated the luminosity for the new GAIA distance to be $13.1\pm3.0~\rm{L}_\odot$. From the bolometric luminosity and effective temperature, we derive a stellar radius of $R_* = 2.1\pm0.2~\rm{R}_\odot$. We use {\sc bagemass} \citep{Maxted2015}, a stellar evolution MCMC code, to determine the stellar mass and age from the effective temperature, metallicity and stellar density\footnote{The stellar density is obtained from the fit to the transit light curve as described in Sect. \ref{sec:photfit} using the relations of \citet{Seager2003}.}, finding $1.72\pm0.07~\rm{M}_\odot$ and $1.0\pm0.2~\rm{Gyr}$, respectively. As an extra check we also compute the stellar radius from the {\sc bagemass} output and find a value of $1.97 \pm 0.07~\rm{R}_\odot$, consistent with the radius derived from the GAIA parallax.

From the HERMES radial velocity pipeline\footnote{See footnote \ref{note:hermes}} we also obtain the stellar equatorial rotation speed $v\sin i_\star = 106.7\pm0.4~\rm{km~s}^{-1}$.
 
\begin{table*}
\centering
\caption{Transit model parameters and their best-fit values obtained from fits to the MASCARA data, NITES data, and the combined photometric data set.}
\begin{tabular}{l c c c c c c c}
 Parameter & Symbol & Units & MASCARA & NITES & MASCARA + NITES\\
 \hline
 \hline
 Reduced chi-square & $\chi^2_\nu$ & - & $1.98$ & $0.51$ & $1.98$ \\
 Norm. MASCARA & $F_0$ & - & $1.00021\pm0.00004$ & - & $1.00032\pm0.00004$ \\ 
 Norm. NITES & $F_1$ & - & - & $1.0000\pm 0.0002$ & $0.99962\pm0.00009$ \\
 Norm. NITES & $F_2$ & - & - & $1.0002\pm 0.0002$ & $0.99975\pm0.00009$ \\
 Epoch & $T_p$ & BJD &$2457097.277\pm0.002$ & $2457097.2751\pm0.0009$ & $2457097.278\pm0.002$ \\  
 Period & $P$ & d & $2.14879\pm0.00001$ & 2.14879 (fixed) & $2.148780\pm0.000008$ \\
 Duration & $T_{14}$ & h & $4.06^{+0.08}_{-0.07}$ & $4.1^{+0.08}_{-0.06}$ & $4.05\pm0.03$ \\
 Planet-to-star ratio\tablefootmark{a} & $p=R_p/R_*$ & - & $0.063^{+0.002}_{-0.001}$ & $0.078^{+0.002}_{-0.001}$ & $0.0735\pm0.0007$ \\
 Impact parameter & $b$ & - & $0.2\pm0.2$ & $0.3\pm0.2$ & $0.2\pm0.1$ \\ 
 Eccentricity & $e$ & - & $0$ (fixed) & $0$ (fixed) & $0$ (fixed) & \\
\end{tabular}
\tablefoot{
\tablefoottext{a}{The best fit value of $p$ differs significantly between fits and as we are unable to reconcile the different data we adopt $p=0.07\pm0.01$ as our final value.}  
}
\label{tab:photpars}
\end{table*} 
 
\section{System parameters}
\label{sec:system}

System parameters are derived from modelling the available photometric and spectroscopic observations. The transit shape is fitted using the available MASCARA and NITES photometric data. The mass of the planet is constrained using the HERMES radial velocity measurements, and the projected spin-orbit angle is obtained from the SONG transit data. 

\subsection{Photometric transit fit} 
\label{sec:photfit}

A \citet{Mandel2002} model is fitted to the photometric data taken by MASCARA using a Markov-chain Monte Carlo (MCMC) approach with the {\sc Python} packages {\sc batman} and {\sc emcee} \citep{Kreidberg2015,FM2013}. We fit a circular transit model, optimizing for the normalization $F_0$, the transit epoch $T_p$, the orbital period $P$, the transit duration $T_{14}$, the planet-to-star radius ratio $p$ and the impact parameter $b$ using uniform priors on all parameters. We employ a quadratic limb darkening law with fixed coefficients from \citet{Claret2000}, using values of $u_1=0.2609$, $u_2=0.3405$, appropriate for the host star. Subsequently, the NITES photometry is also fitted using the same model, except with the orbital period fixed to the best fit value obtained from the MASCARA photometry and separate normalization factors $F_1$ and $F_2$ for each transit. In the modelling presented here we do not include the possible effects of non-spherical geometry, which can influence $T_{14}$ and $b$, or gravity darkening, which can modify $p$. Both of these effect might play a role for a rapid rotator such as MASCARA-1. The best-fit parameter values and uncertainties, obtained from the median and the 16th and 84th percentiles of the output MCMC chains, are listed in Table \ref{tab:photpars}. 

We find that all orbital parameters are consistent between the MASCARA and NITES data with exception of the planet-to-star ratio $p$, which takes the value of $0.063^{0.002}_{-0.001}$ for the MASCARA data and $0.078^{0.002}_{-0.001}$ for the NITES data. We consider the possibility that this difference in the value of $p$ is a result of blended light in the MASCARA photometry, caused by the $1\arcmin$ pixel scale of MASCARA,  which would decrease the transit depth and thus the value of $p$. However, a simple calculation shows that we would need ${\sim}35\%$ blended light in the MASCARA aperture which is more than can be explained by the background sources present, which we conservatively estimate to contribute a maximum of ${\sim}20\%$ of the total light. Another possibility comes from the algorithm used to process the MASCARA data, as we are still in the process of investigating possible systematic changes introduced by our methods.

To further investigate this discrepancy we perform a joint fit to the MASCARA and NITES photometry. The reduced chi-square value for the NITES fit indicates that we are over-estimating the uncertainties on the photometry, this is plausible since we assume a conservative model for the scintillation noise, the dominant source of noise in the observations, when calculating the uncertainties. To ensure we place similar weight on both the MASCARA and NITES observations we reduce the NITES uncertainties by a factor 2 for the joint fit. The results of the joint fit are shown in Fig. \ref{fig:photometry} and the parameters are listed in Table \ref{tab:photpars}. From the joint fit we obtain $p=0.0735\pm0.0007$, significantly different from both the NITES and MASCARA values. It should also be noted that the lack of baseline in the NITES observations results in a lower normalization in the joint fit, helping to reduce any dicrepancy in the depths. We are unable at this time to explain this discrepancy and adopt a value of $p=0.07\pm0.01$, midway in between and consistent with both the MASCARA and NITES values, for the derivation of the planet radius listed in Table \ref{tab:syspars}. 

\subsection{Radial velocities and planet mass}

\begin{figure}
  \centering
  \includegraphics[width=8.5cm]{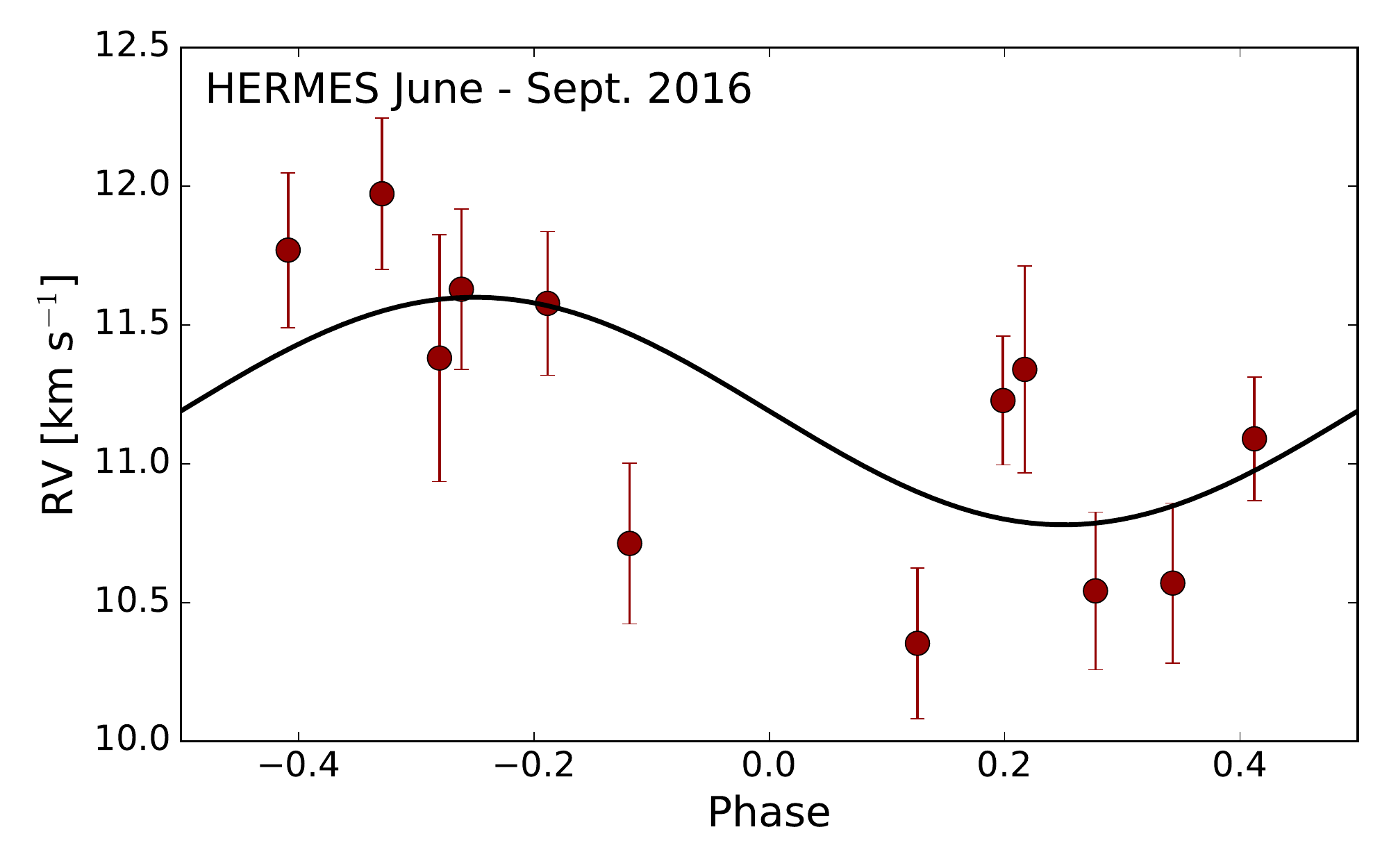}
  \caption{HERMES RV measurements of MASCARA-1 obtained from July to September 2016. Shown are the data (red points) and the best-fit model (black line). The high projected rotation speed of the star, $v\sin i_\star = 106.7~\rm{km~s}^{-1}$, results in relatively large uncertainties on the RV measurements of order $300\rm{~m~s}^{-1}$.}
  \label{fig:RV}
\end{figure}

Stellar radial velocities and their uncertainties are determined from the HERMES spectra using the standard pipeline\footnote{See footnote \ref{note:hermes}}, in which a mask of individual lines is cross-correlated with the data to obtain an average line profile. For the observations of MASCARA-1 the {\sc HermesF0Mask} was used, being the closest to the spectral type of the star of the available masks. The average line profile is subsequently fitted with a rotationally broadened line profile, the dominant source of broadening, to obtain the radial velocities and their uncertainties. The obtained uncertainties are relatively large (${\sim}300\rm{~m~s}^{-1}$) due the high equatorial rotation speed of the star. The epochs of the radial-velocity data points are phase-folded using the best-fit orbital solution derived from the photometry. A circular orbital solution was fitted to the data keeping the period and epoch fixed to the values derived above. The RV data and best-fit solution are shown in Fig. \ref{fig:RV}, and the best-fit parameters are listed in Table \ref{tab:specpars}. The amplitude of the stellar radial velocity variation is not well constrained to $K=400\pm100~\rm{m~s}^{-1}$, corresponding to a planet mass of $M_p = 3.7\pm0.9~\rm{M}_{\rm{jup}}$.
 
\subsection{Rossiter-McLaughlin effect and projected spin-orbit angle}

To consolidate the confirmation of the planetary nature of the companion, and to determine the projection of the spin-orbit angle (obliquity), we observed the systems during a transit with the SONG telescope (See section \ref{sec:observations}). Simple Cross Correlation Functions (CCFs) are created from the obtained spectra using a stellar template appropriate for MASCARA-1. \footnote{The template was created using the interactive data language (IDL) interface SYNPLOT (I. Hubeny, private communication) to the spectrum synthesis program SYNSPEC \citep{Hubeny1995}, utilizing a Kurucz model atmosphere (see \url{http://kurucz.harvard.edu/grids.html}).} The CCFs are shown in Fig. 3 panel A (top left).

To analyze the distortions of the stellar absorption lines and measure the projected obliquity ($\lambda$) we use the code presented in \citet{Albrecht2007}, including updates presented in \citet{Albrecht2013}. In short, we create a pixelated stellar disk where each pixel is assigned a radial velocity based on stellar rotation, and micro- and macro-turbulence \citep[see e.g.][]{Gray1984}. Here the Point Spread Function (PSF) of the spectrograph is included in the width of the Gaussian function representing the micro-turbulence. We do not include convective blueshift as the CCFs do not have a high enough SNR to constrain this parameter and in turn the influence on $\lambda$ is small. The light contribution of each pixel in our model is governed by a quadratic limb darkening model. For each observation, a model stellar absorption line is created. During transit, the integration is carried out only over the surface not blocked by the companion. These model CCFs are then compared to the observed CCFs, taking into account shifts due to the orbital velocity. 

\begin{table*}
\centering
\caption{Model parameters used in fitting the RV and RM measurements and their best-fit values.}
\begin{tabular}{l c c c c c c c}
 Parameter & Symbol & Units & HERMES & SONG \\
 \hline
 \hline
 Epoch\tablefootmark{a} & $T_p$ & BJD  & $2457097.278$ (fixed) & $2457097.283\pm0.0022$\\  
 Period & $P$ & d & $2.148780$ (fixed) &  $2.148780$ (fixed) \\
 Duration\tablefootmark{a} & $T_{14}$ & h & - & $4.15\pm0.0030$\\
 Planet-to-star ratio\tablefootmark{a} & $p=R_p/R_*$ & - & - & $0.0795\pm0.00065$ \\
 Impact parameter\tablefootmark{a} & $b$ & - &   - & $0.122\pm 0.012$ \\ 
 Eccentricity & $e$ & - & $0$ (fixed) & $0$ (fixed)\\
 RV amplitude\tablefootmark{a} & $K$ & m s$^{-1}$ & $400\pm100$ & $190\pm90$\\
 Systemic velocity\tablefootmark{b} & $\gamma$  & km s$^{-1}$ & $11.20\pm0.08$ & $8.52\pm0.02$\\
 Projected obliquity & $\lambda$ & \degr & - & $69.5\pm4$\\ 
 Micro turbulence & $\nu$ & km s$^{-1}$ & - & $0.4 \pm 0.3$ \\
 Macro turbulence & $\zeta$ & km s$^{-1}$ & - & $7.3 \pm 0.2$ \\
 Projected rotation speed & $v \sin i_\star$ & km s$^{-1}$ & - & $109\pm3$\\
\end{tabular}
\tablefoot{
\tablefoottext{a}{For the SONG analysis, the best-fit parameters from the photometry and radial velocities were used to set Gaussian priors on $T_p$, $T_{14}$, $p$, $b$ and $K$.}
\tablefoottext{b}{We find a significant offset between the values for $\gamma$ derived from the HERMES and SONG data. An analysis of our methods revealed the offset likely originates from a difference in the spectral templates used in reducing the HERMES and SONG data, but does not influence our results.}
}
\label{tab:specpars}
\end{table*}

\begin{figure*}
  \centering
  \includegraphics[width=8.5cm]{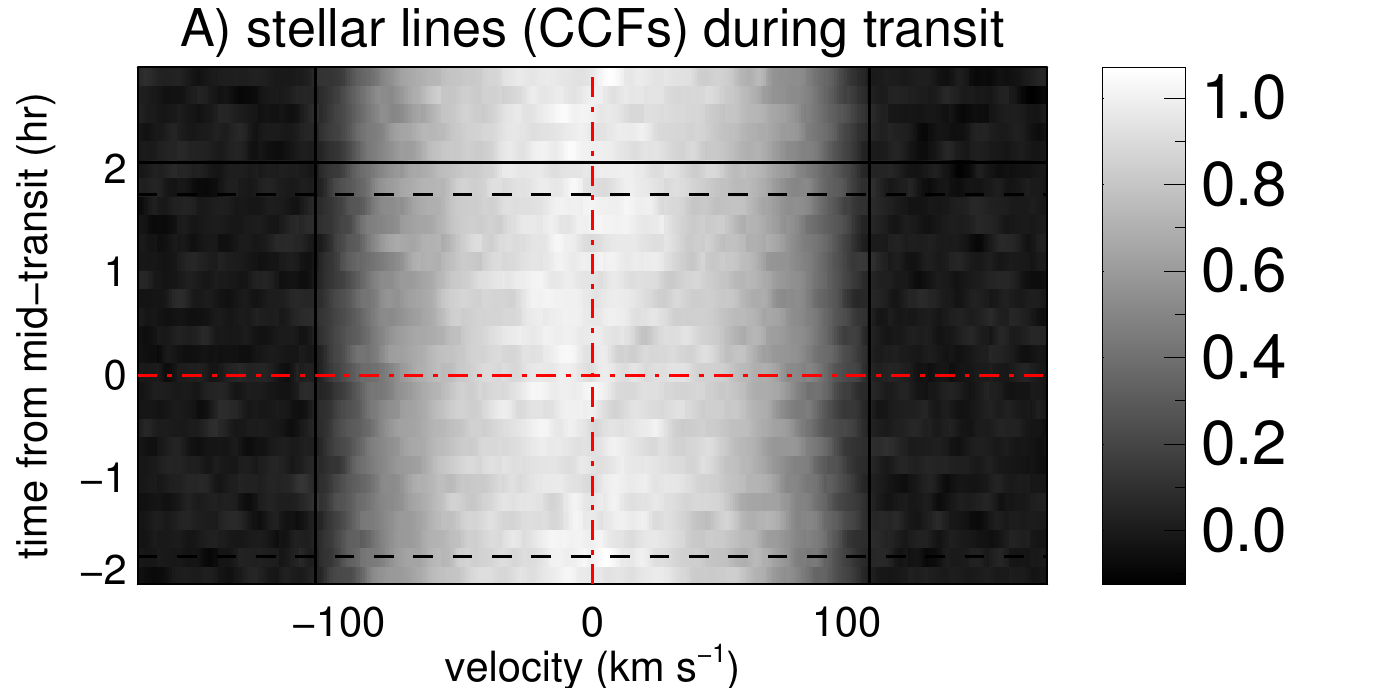}
  \includegraphics[width=8.5cm]{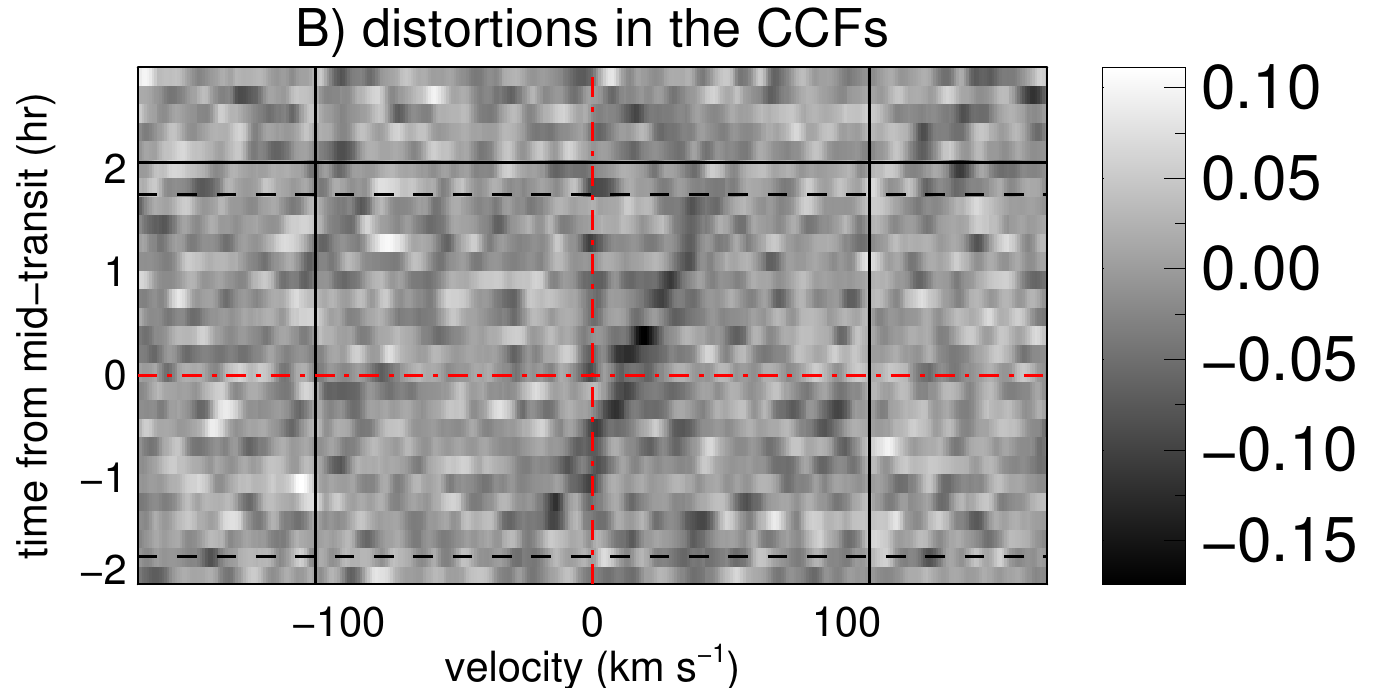}
  \includegraphics[width=8.5cm]{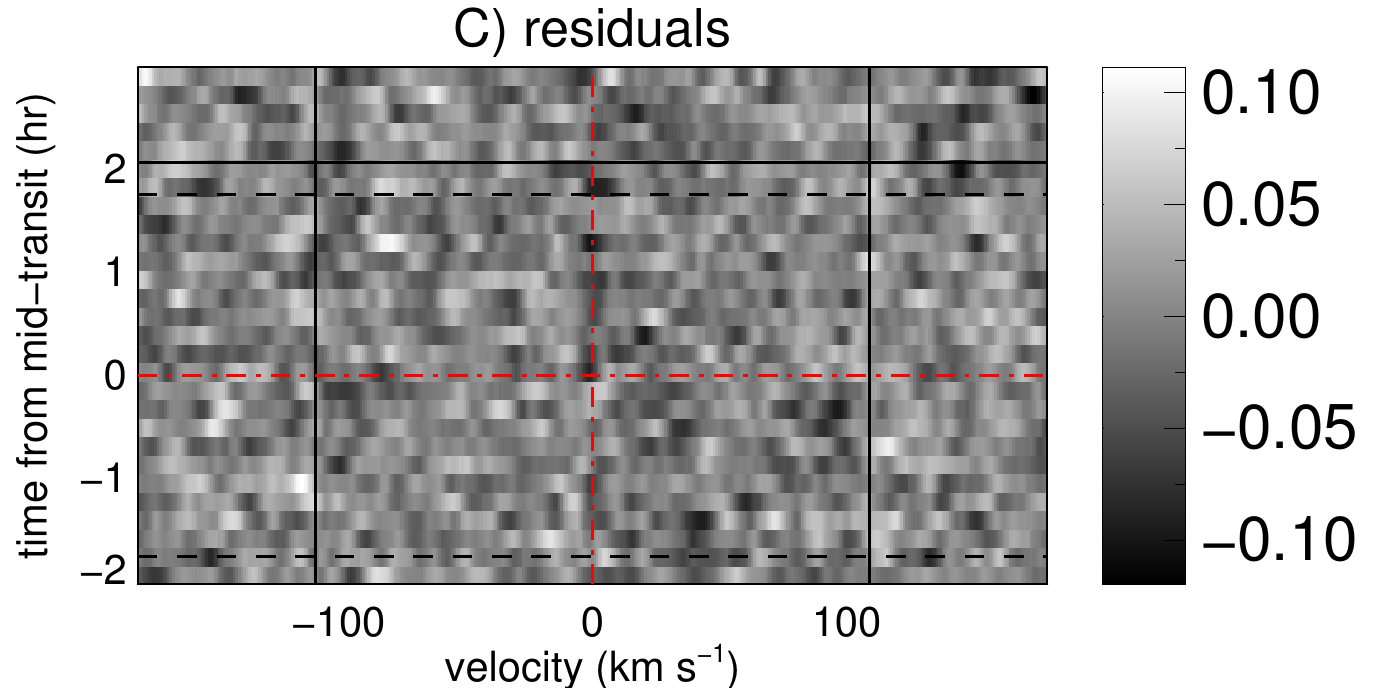}
  \includegraphics[width=8.5cm]{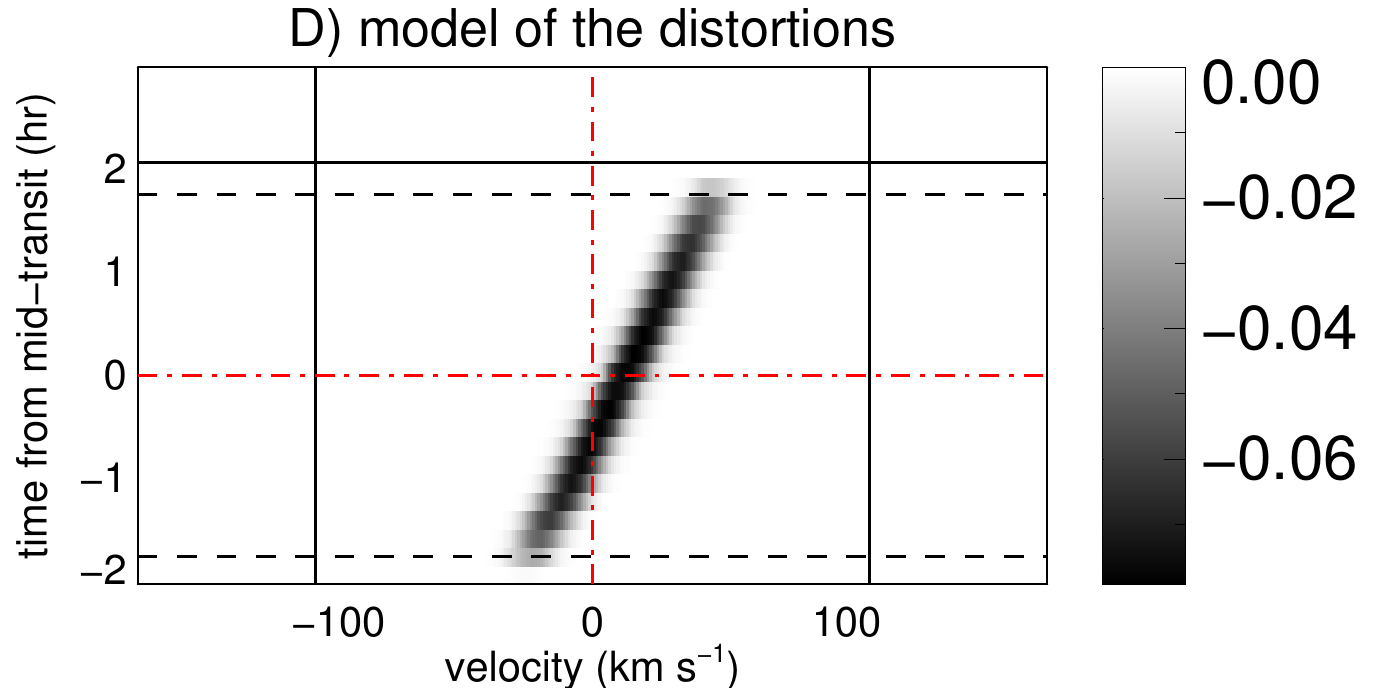}
  \caption{Observations of MASCARA-1 during and after the transit of MASCARA-1\,b. In each panel, the solid horizontal line indicates the end of transit, and the dashed lines the end of ingress and beginning of egress. The solid black vertical lines show the best-fit value for $v \sin i_\star$. The dashed-dotted red lines indicate the mid-transit time and the rest-frame velocity of the star. 
{\it Panel A:} The observed CCFs in gray scale. {\it Panel B:} The same as in Panel A, but with the best-fit model for the undisturbed stellar CCF subtracted, isolating the distortions due to the planetary transit over the rotating stellar photosphere. {\it Panel C:} Residuals after our best fitting model -- including the planetary transit -- is subtracted. {\it Panel D:} The corresponding model to the data shown in Panel B.}
  \label{fig:RM}
\end{figure*}

We run an MCMC to obtain confidence intervals for $\lambda$ and $v \sin i_\star$, the parameters of interest.  For this we allowed the following parameters to vary: $\lambda$, $v \sin i_\star$, macro-turbulence ($\zeta$), micro-turbulence ($\nu$), the transit epoch $T_{p}$, planet-to-star radius ratio $p$, scaled semi-major axis $a/R_*$, the cosine of the orbital inclination $\cos i$, the stellar systemic velocity $\gamma_{\rm SONG}$, the stellar RV amplitude  $K$, and limb darkening parameters $u_1+u_2$. We use prior information from the photometric and radial-velocity observations for the parameters $T_{p}$, $a/R_*$, $p$ and $K$. We further used the prior information for the impact parameter $b$, and the transit duration $T_{14}$ also derived from photometry (Table~\ref{tab:photpars}), $0.6\pm0.1$ for $u_1+u_2$ \citep{Claret2013}, and uniform priors on all other parameters. Finally for each calculation of the likelihood we allow each of the observed CCFs to be offset and scaled in intensity \citep[see also][]{Albrecht2013}. This way we include the influence of the none perfect normalization of the spectra into the uncertainty interval of the final parameters. A mismatch in the continuum normalization of the observed spectrum and the template spectrum causes an offset in the ``baseline'' of the obtained CCF.  In addition, the size of the CCFs depends on the SNR of the spectra, which did vary throughout the observations. In Tab.~\ref{tab:specpars} we report the results for the above parameters, including the $68\%$ confidence intervals from the analysis of the CCFs.

With an projected obliquity of $69.5\pm0.3^\circ$ we find that the rotation axis of MASCARA-1 is misaligned with respect to the orbit of the planet. This can also be seen in Fig.~\ref{fig:RM} panel B, where the planet shadow is isolated. Despite a low impact parameter the trail of the planet does not reach from $-v \sin i_\star$ to $+v \sin i_\star$, as it would be the case if the planet would travel near the stellar equator. For central transits and large misalignments the amplitude of the RM effect is reduced, as the planet travels from pole to pole, and not over the equator. If the impact parameter would have been larger, MASCARA-1\,b would have traveled largely over the receding stellar surface area. We also find that the SONG data prefer a small impact parameter ($b=0.122\pm0.012$), consistent with the photometry ($b=0.2\pm0.1$). The transit mid-point is found 7~min later than expected from the photometric ephemeris.  We argue that the formal uncertainty in the  projected rotation speed ($v \sin i_\star=109.0\pm0.1$~km\,s$^{-1}$) underestimates the true uncertainty in this parameter. MASCARA-1 is a fast rotator and we expect a departure from the perfect spherical shape assumed in our analysis. In addition, we also expect a significant gravity darkening. This would result in a dark band along the equator for large $i_\star$ and an apparent increase in the stellar limb darkening for low $i_\star$. We therefore argue that an uncertainty of 3\,km\,s$^{-1}$ is more appropriate for the projected rotation speed.  We find that our data prefers a larger than expected limb darkening  $u_1+u_2 =0.9\pm0.1$ which we attribute to the simplified physics in our model rather than to a true disagreement with the model limb darkening parameters. Indeed if our data would be of high SNR, we might be able to determine $i_\star$ and therefore the actual obliquity ($\Psi$) in this system from the RM data alone. However given the SNR in our SONG data we postpone such an analysis until such a dataset is available. We also expect the uncertainty in $\lambda$ to be about $4^\circ$, larger then the formal value quoted above, for the same reasons that we expect the $v \sin i_\star$ uncertainty to be underestimated.

\section{Discussion and Conclusions}
\label{sec:discussion}

\begin{figure}
  \centering
  \includegraphics[width=8.5cm]{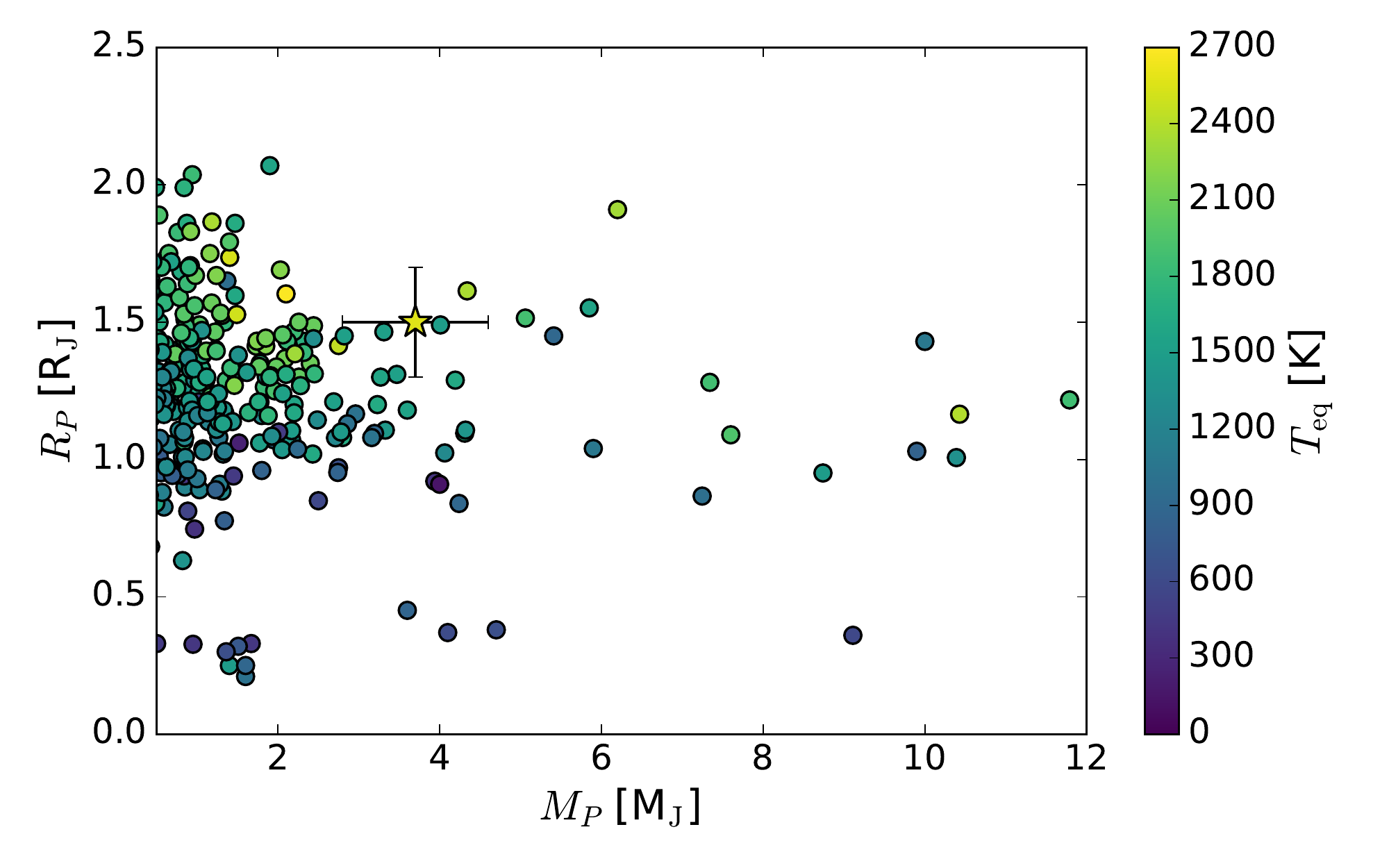}
  \caption{Mass-radius relationship for transiting exoplanets. MASCARA-1\,b is indicated by the star marker and the points are coloured by the theoretical equilibrium temperature.}
  \label{fig:mass_radius}
\end{figure}

\begin{figure}
  \centering
  \includegraphics[width=8.5cm]{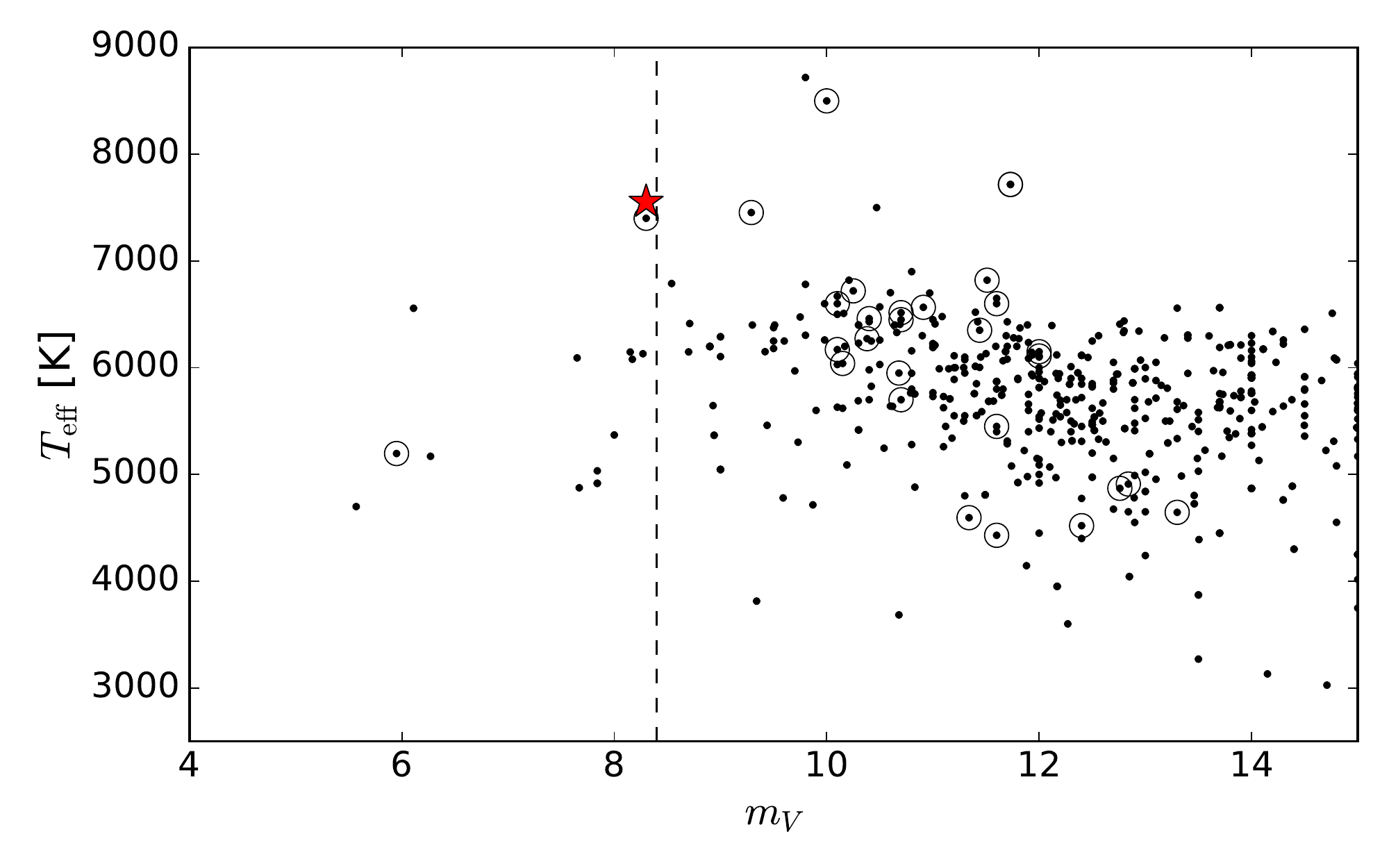}
  \caption{Effective temperature versus visual magnitude for stars hosting transiting exoplanets. MASCARA-1 is indicated by the red star and systems for which the obliquity has been measured are indicated by open circles. MASCARA targets stars to the left of the dashed black line ($4<m_V<8.4$).}
  \label{fig:hosts}
\end{figure}

\begin{table}
\small
\centering
\caption{Parameters describing the MASCARA-1 system, derived from the best-fit models to the photometric and spectroscopic data.}
\begin{tabular}{l c c c}
 Parameter & Symbol & Value \\
 \hline
 \hline
 Stellar parameters \\
 \hline
 Identifiers & & HD\,201585, HIP\,104513\\
 Right Ascension & & $21^h10^m12.37^s$ \\
 Declination & & $+10\degr44\arcmin19.9\arcsec$ \\
 Spectral Type & & A8\\
 V-band magnitude & $m_V$ & 8.3 \\
 Age\tablefootmark{a} & & $1.0\pm0.2~\rm{Gyr}$\\
 Effective temperature\tablefootmark{b} & $T_{\rm{eff}}$ &  $7554\pm150~\rm{K}$ \\
 Projected rotation speed & $v\sin i_\star$ & $109\pm4\rm{~km~s}^{-1}$ \\
 Surface gravity & $\log g$ & 4\\
 Metallicity & [Fe/H] & 0\\  
 Stellar mass\tablefootmark{a} & $M_*$ & $1.72\pm0.07~\rm{M}_\odot$ \\
 Stellar radius & $R_*$ & $2.1\pm{0.2}~\rm{R}_\odot$ \\
 Stellar density & $\rho_*$ & $0.33^{+0.02}_{-0.04}~\rm{g~cm}^{-3}$ \\
 \hline
 Planet parameters\\
 \hline
 Planet radius\tablefootmark{c} & $R_p$ & $1.5\pm0.3~\rm{R}_{\rm{Jup}}$ \\
 Planet mass & $M_p$ & $3.7\pm0.9~\rm{M}_{\rm{Jup}}$ \\
 Planet density & $\rho_p$ & $1.5\pm0.9~\rm{g~cm}^{-3}$\\
 Equilibrium temperature\tablefootmark{d} & $T_{\rm{eq}}$ & $2570^{+50}_{-30}~\rm{K}$\\
 \hline
 System parameters\\
 \hline  
 Epoch & $T_p$ & $2457097.278\pm0.002$ BJD \\
 Period & $P$ & $2.148780\pm0.000008$ days \\
 Semi-major axis & $a$ & $0.043\pm0.005~\rm{AU}$ \\
 Inclination & $i$ & $87\degr{}^{+2}_{-3}$ \\
 Eccentricity & $e$ & 0 (fixed) \\
 Projected obliquity & $\lambda$ & $69.5\degr\pm3$ \\
 \hline
\end{tabular}
\tablefoot{
\tablefoottext{a}{Computed using the {\sc BAGEMASS} code from \citet{Maxted2015}}
\tablefoottext{b}{Taken from \citet{McDonald2012}, assuming a typical uncertainty.}
\tablefoottext{c}{Assuming $p=0.07\pm0.01$.}
\tablefoottext{d}{Computed assuming uniform redistribution and a Bond albedo of zero.}
}
\label{tab:syspars}
\end{table} 

Table \ref{tab:syspars} lists the final physical parameters describing the MASCARA-1 system. We find a planetary mass and radius of $3.7\pm0.9~\rm{M}_{\rm{Jup}}$ and $1.5\pm0.3~\rm{R}_{\rm{Jup}}$, albeit with a large uncertainty on the mass caused by the high spin-rotation velocity of the star. The uncertainty on $M_p$ can be reduced by obtaining more RV measurements and we have started a monitoring campaign for this purpose. MASCARA-1\,b orbits a bright A star in $2.148780\pm8\times10^{-6}$ days at a distance of $0.043\pm0.005~\rm{AU}$. The high temperature of the host star means that MASCARA-1\,b has a high equilibrium temperature of $2570^{+50}_{-30}~\rm{K}$ ($A_B=0$), making it one of the hottest gas giants known.

Figure \ref{fig:mass_radius} shows the location of MASCARA-1\,b in the planetary mass-radius diagram. From this we can see that while the radius is large it is not the most extreme case found to date, though it is the most-irradiated of the more massive ($M_p > 3~\rm{M}_{\rm{Jup}}$) hot Jupiters. MASCARA-1\,b also follows the empirical relationship between mass, radius, equilibrium temperature, host star metallicity and tidal heating from \citet{Enoch2012}. Figure \ref{fig:hosts} shows the location of MASCARA-1 compared to other stars hosting transiting exoplanets as a function of visual magnitude and effective temperature. It is clear that MASCARA-1 lies in a part of host star parameter space that has been largely unexplored to date, being a bright early-type star. 

Currently there are only a few host stars with effective temperatures larger than $7000$~K for which the stellar obliquity is known (e.g. WASP-33 \cite{CollierCameron2010}, KELT-17 \cite{Zhou2016}, Kepler-13 \cite{Mazeh2012}, see also Fig. \ref{fig:hosts})\footnote{Visit \url{www.astro.keele.ac.uk/jkt/tepcat/rossiter.html} for an up-to-date list of obliquity measurements.} while some constraints exists for other similar systems such as HAT-P-57 \citep{Hartman2015}. This small group of systems does display a large spread in spin orbit alignments and MASCARA-1 does confirm this trend. So far it is not clear what causes the large obliquities. They might be caused by dynamical interactions \citep[e.g. ][]{Fabrycky2007,Nagasawa2008}. However they might also be a general feature of star formation. Proposed mechanisms include: chaotic star formation \citep{Bate2010,Thies2011}, magnetic star-disk interaction \citep{Lai2011,Foucart2011,Spalding2015}, torques due to neighbouring stars \citep{Batygin2012,Lai2014}, tidal dissipation \citep{Rogers2013a}, and internal gravity waves \citep{Rogers2013b}. We can differentiate between the different theories by measuring obliquities in systems with varying multiplicity, planet mass, orbital separation, and stellar mass and structure \citep[e.g.][]{Albrecht2013}. Detecting and characterizing exoplanets orbiting early type main sequence stars should lead to a better understanding of the environment in which planets form.
 
Since it orbits a bright host star MASCARA-1\,b is a particularly interesting target for atmospheric characterization via transmission spectroscopy. It is reminiscent of the WASP-33 system \citep{CollierCameron2010}, in which a $2.1~M_{\rm{Jup}}$ hot Jupiter transits a $T_{\rm{eff}}=7400~\rm{K}$ A-star. However, while follow-up observations of WASP-33 are significantly hampered by the delta-scuti variations of its host star, no such variability is detected in MASCARA-1. Assuming a temperature of $2550~\rm{K}$ and a hydrogen/helium atmosphere, MASCARA-1\,b will have an atmospheric scale height of $H=215~\rm{km}$, implying that an absorption feature which extends out to $5\times H$ will result in a transmission signal of $0.01\%$. Interestingly, the sodium feature of the host star is significantly weaker than for a solar type star, in addition to being velocity broadened to $106~\rm{km~s}^{-1}$. This will make it significantly more straightforward to isolate the planet atmospheric sodium absorption from stellar effects – e.g due to the Rossiter-McLaughlin effect \citep[e.g.][]{Snellen2008,DiGloria2015}. 

The detection of MASCARA-1\,b shows that the Multi-site All-Sky CAmeRA has the potential to increase the number of transiting exoplanets suitable for high-resolution atmospheric studies \citep[e.g.][]{Snellen2010,Brogi2012} as well as expand our knowledge of planets orbiting early-type stars. The Northern MASCARA station on La Palma has now gathered over 2 years of data and the Southern station in La Silla starts observations in June 2017. We expect to find several more planets around both early- and late-type stars in the coming years. 

\begin{acknowledgements}
IS acknowledges support from a NWO VICI grant (639.043.107). This project has received funding from the European Research Council (ERC) under the European Union's Horizon 2020 research and innovation programme (grant agreement nr. 694513). Based on observations made with the Mercator Telescope, operated on the island of La Palma by the Flemmish Community, at the Spanish Observatorio del Roque de los Muchachos of the Instituto de Astrofísica de Canarias. Based on observations obtained with the HERMES spectrograph, which is supported by the Research Foundation - Flanders (FWO), Belgium, the Research Council of KU Leuven, Belgium, the Fonds National de la Recherche Scientifique (F.R.S.-FNRS), Belgium, the Royal Observatory of Belgium, the Observatoire de Genève, Switzerland and the Thüringer Landessternwarte Tautenburg, Germany. Based on observations made with the Hertzsprung SONG telescope operated on the island of Tenerife by the Aarhus and Copenhagen Universities in the Spanish Observatorio del Teide of the Instituto de Astrofísica de Canarias. The Hertzsprung SONG telescope is funded by the Danish National Research Foundation, Villum Foundation, and Carlsberg Foundation. This research has made use of the SIMBAD database,
operated at CDS, Strasbourg, France. This research has made use of the VizieR catalogue access tool, CDS, Strasbourg, France. We have benefited greatly from the publicly available programming language {\sc Python}, including the {\sc numpy, matplotlib, pyfits, scipy} and {\sc h5py} packages.
\end{acknowledgements}

\bibliographystyle{aa}
\bibliography{../mascara.bib}

\end{document}